\documentclass[twocolumn,showpacs,aps,prb]{revtex4}

\usepackage{natbib}
\usepackage{graphicx}
\usepackage{amsmath,amssymb}
\usepackage{mathrsfs}
\usepackage{subfigure}
\usepackage{boxedminipage}

\newcommand{\avg}[1]{\left\langle #1 \right\rangle}
\newcommand{\redavg}[1]
	{\left\langle\hspace{-0.7mm}\left\langle #1 \right\rangle\hspace{-0.7mm}\right\rangle}
\newcommand{\redavgSm}[1]
	{\langle\hspace{-0.5mm}\langle #1 \rangle\hspace{-0.5mm}\rangle}	
\newcommand{\comm}[1]{\left[#1\right]}

\newcommand{\eq}[1]{Eq.~(\ref{#1})}
\newcommand{\twoEq}[2]{Eqs.~(\ref{#1}) and (\ref{#2})}

\newcommand{\fig}[1]{Fig.~\ref{#1}}

\def \be{\begin{equation}}
\def \ee{\end{equation}}
\def \bmlett{\begin{mathletters}}
\def \emlett{\end{mathletters}}

\def \LL{{\mathcal L}}

\def \hrho {\hat{\rho}}

\def \fm {\chi}
\def \dm {\hat{\rho}}

\def \xop {\hat{x}}
\def \pop {\hat{p}}
\def \mop {\hat{m}}
\def \tunx {\tau_1}
\def \tuno {\tau_0}
\def \phase {\phi}
\def \xx {\bar{x}}
\def \pp {\bar{p}}
\def \var {V}
\def \xvar {\var_x}
\def \pvar {\var_p}
\def \xpvar {\var_{xp}}
\def \zx {\Delta x_0}
\def \zp {\Delta p_0}
\def \Hleads {H_{leads}}

\def \charFunc {\Phi}

\def \Tcoth {{\tilde{T}_0}}

\def \tstar {t^*}
\def \smallfigsize {0.48\textwidth}
\def \bigfigsize{0.4\textwidth}

\def \Gx {\tilde{\Gamma}(x)}

\begin{document}

\title{Full counting statistics and conditional evolution
in a  nanoelectromechanical system}
\author{S. D. Bennett and A. A. Clerk}
\date{\today}
\affiliation{Department of Physics, McGill University, Montr\'{e}al, Qu\'{e}bec, Canada H3A 2T8}

\begin{abstract}
We study theoretically the full distribution of transferred 
charge in a tunnel junction (or quantum point contact) 
coupled to a nanomechanical oscillator,
as well as the conditional evolution of the oscillator. 
Even if the oscillator is very weakly coupled to
the tunnel junction, it
can strongly affect the tunneling statistics and
lead to a highly non-Gaussian distribution.
Conversely, given a particular measurement history of the current, the
oscillator energy distribution may be localized and highly non-thermal.
We also discuss non-Gaussian correlations between the oscillator
motion and tunneling electrons; 
these 
show that the tunneling back-action cannot
be fully described 
as an effective thermal bath coupled to the oscillator.
\end{abstract}

\maketitle

\section{Introduction}

The possibility to observe the quantum mechanics
of a macroscopic object
has sparked
significant  interest in
nanoelectromechanical systems (NEMS), which consist of
a mechanical oscillator coupled to a mesoscopic conductor.
In recent experiments,
the oscillator motion has been measured
with near quantum-limited precision
using the conductor as a detector
\cite{Flowers-Jacobs07,Poggio08,Knobel03},
and cooling of the oscillator by quantum back-action
has been observed \cite{Naik06}.
In these experiments it is the current noise of the conductor (i.e.~the second moment of
current fluctuations)
that is used 
to measure position fluctuations of the oscillator.
The effect of the oscillator on the current noise has been well studied
theoretically, leading to new understanding of back-action
and quantum dissipation in NEMS 
\cite{Armour04,ZhangBlencowe02,Usmani07,ClerkGirvin04,Mozyrsky02,Doiron08,Wabnig05}.
However, much more information
lies in the {\it full} probability distribution of transmitted charge through the conductor, 
or the full counting statistics (FCS)
\cite{Levitov04}.
In addition to being of theoretical interest, FCS is
an experimentally accessible quantity
and the third moment was recently measured 
in a tunnel junction
\cite{Bomze05}.
Still more information may be gained by considering 
the conditional evolution: 
given a particular measurement history for the current,
what can we say about the state of the oscillator?

In this paper, we
study the full statistics of a tunnel junction (or quantum point contact) coupled to a 
nanomechanical oscillator, a system
recently realized in experiment \cite{Flowers-Jacobs07,Poggio08}.
This system is a prime candidate for measuring FCS in NEMS, since the intrinsic
shot noise can dominate over other noise sources making it feasible
to measure the higher moments.
Starting from a microscopic, fully quantum model,  we calculate
the FCS of tunneled charge as well as
the conditional evolution of the oscillator,
and find several surprises that would 
not be apparent in a study of the noise
alone.  
Despite weak oscillator-conductor coupling, 
we find that the oscillator can 
strongly enhance the third and higher moments 
of the FCS,
leading to a markedly non-Gaussian distribution.
This results from long-lived energy fluctuations in the high-$Q$ oscillator, which
allow correlations between the oscillator motion and tunneling electrons to accumulate up
to the ringdown time of the oscillator, overwhelming the weak coupling strength and
dominating the FCS.
Further, even though the conductor couples linearly to the oscillator position, the oscillator state conditioned on a particular measurement of current can be highly localized in energy.
Finally, we discuss non-Gaussian 
correlations between the current and back-action force
on the oscillator that are not captured by treating the tunnel junction
as an effective thermal bath.
These correlations arise
from the random momentum kicks imparted
to the oscillator by individual tunneling electrons,
which cause one half of the back-action to be
correlated with tunneling.
The non-Gaussian correlations 
lead to 
signatures in measurable quantities such as the
current noise; thus,
measuring the current noise could provide
a way to distinguish quantum back-action 
from the effects of an equilibrium bath.

Note that FCS were studied previously in a very different type of NEMS, a charge shuttle 
\cite{Pistolesi04,Flindt05}.  Conditional evolution in NEMS were studied
using a quantum optics approach \cite{Ruskov04,Hopkins03}, but these studies miss key features arising in our microscopically-derived model.  The average current and current 
noise of the 
NEMS studied here were addressed in 
Refs.~\cite{ClerkGirvin04,Doiron08,Mozyrsky02,Wabnig05}; 
unlike these works, we present 
an exact solution of the master equation and  study the FCS.

\section{Master equation and its solution}

The Hamiltonian of the coupled system is $H = H_{osc} + \Hleads + H_T$,
where
$H_{osc}$ describes a harmonic oscillator of mass $M$ and frequency $\Omega$
including dissipation due to an equilibrium thermal bath at temperature $T_0$ \cite{Caldeira83}.
Electrons in the leads are described by $\Hleads=  \sum_{\alpha,k} \varepsilon_k c_{\alpha k}^\dag c_{\alpha k} - eV \mop$,
where $c_{\alpha k}$  annihilates an electron in lead $\alpha=L,R $, $V$ is
the junction bias voltage, 
and the operator $\mop$ counts the number of tunneled electrons. 
$H_T$ describes 
electron tunneling for the experimentally relevant
case of weak oscillator-junction coupling
\cite{Mozyrsky02,ClerkGirvin04,Doiron08,Wabnig05},
\begin{equation}
     H_T  =  \frac{\tuno +  e^{i\eta} \tunx \xop}{2\pi\Lambda}
    \sum_{k,k'} \left( Y^\dag c_{Rk}^\dag c_{Lk'} + h.c.  \right),
\label{eq:Ht}
\end{equation}
where $\Lambda$ is the lead density of states, 
$\eta$ describes the dependence of the transmission phase
on the oscillator position $\xop$,
and $Y^\dag$ 
is the raising operator associated with 
$\mop$,  e.g. $\comm{\mop,Y^\dag} = Y^\dag$.
We focus on an inversion symmetric system 
in which $\eta$ vanishes \cite{ClerkGirvin04}.

We describe the system using a reduced density matrix 
$\hrho(t)$
tracking the state of the oscillator and $m$,
the number of tunneled electrons.
In the weak tunneling regime of interest, the off-diagonal (in $m$)
elements of $\dm$ decouple from the diagonal elements.
Since our aim is to calculate the statistics of $m$, we need only
consider the diagonal
elements, $\hrho(m;t) = \avg{m | \hrho(t) | m}$.
Treating $H_T$ perturbatively and making a standard Markov
approximation which requires $eV \gg \hbar \Omega$, 
we derive a master equation for 
$\hrho(m;t)$
\cite{ClerkGirvin04}.
Fourier transforming in $m$,
$\hrho(\fm;t) = \sum_{m=-\infty}^\infty e^{i\fm m}  \hrho(m;t)$,
the  equation reads
\begin{align}
\label{eq:dm2}
   \partial_t  {\hrho}(\fm;t) =  &-\frac{i}{\hbar} \big[ H_0,\hrho \big]
    -  \frac{i \gamma} {2\hbar} 
    \big[ \xop, \left\{\pop,\dm\right\} \big] 
    - \frac{D}{\hbar^2}  \big[ \xop,\left[\xop,\dm\right] \big] 
    \nonumber \\
    &+ \frac{\left(e^{i\fm}-1\right)}{\tunx^2}  
    \bigg[ \frac{2D_1}{\hbar^2}
    \left( \tuno + \tunx \xop \right) \dm
    \left( \tuno + \tunx \xop \right) 
     \\
     &+ \frac{i\gamma_1}{2\hbar}
    \left[ \tuno\tunx \left( \pop\dm-\dm\pop \right)
    + \tunx^2 \left( \pop\dm\xop - \xop\dm\pop \right) \right]
    \bigg].
    \nonumber
\end{align}
Here, $H_0$ describes the coherent dynamics 
of the oscillator, and
the total damping and diffusion coefficients are 
$\gamma = \gamma_0 + \gamma_1$ and $D = D_0+D_1$.
The coefficients $\gamma_0$ and 
$D_0 =  (M \gamma_0 \hbar\Omega/2) \coth{\left( \hbar\Omega / 2 T_0 \right)}$ 
are associated with 
the equilibrium bath ($k_B = 1$),
while $\gamma_1$ and $D_1$ 
describe back-action damping and diffusion
due to the tunnel junction.
Taking 
the electronic temperature in the leads to be much less
than $eV$ \footnote{This limit is in no way required, but yields simple expressions
for the back-action
damping and temperature.}, 
these
are given by
$\gamma_1  = \hbar \tunx^2 / 2\pi M$ and
$D_1 = M \gamma_1 T_1$,
where the effective temperature due to the tunnel junction is
$T_1 = eV/2$  \cite{ClerkGirvin04,Mozyrsky02}.  
Note that if we average over $m$ (i.e. set $\fm=0$), 
\eq{eq:dm2} reduces to the quantum Brownian motion master 
equation for an oscillator coupled to two effective thermal baths consisting of
the environment and the tunnel junction \cite{Caldeira83}.
Conversely, tracing over the oscillator degrees of freedom yields the generating
function for the FCS,
\begin{equation}
\label{eq:fcs}
	\charFunc(\fm;t) = \text{tr} \left[ \dm(\fm;t) \right] = \textstyle\sum_m e^{i \fm m} P(m;t),
\end{equation}
where $P(m;t)$ is the probability that $m$ electrons have tunneled in time
interval $t$.  
Note that the trace of $\hrho(t)$ over {\it all} degrees of freedom is
$\sum_m \text{tr} \left[ \dm(m;t) \right] = 1$.

The above model has been used to study 
the average current and noise
\cite{ClerkGirvin04,Doiron08,Mozyrsky02,Wabnig05};
here we present its exact solution
and use it to study FCS and conditional evolution.
To work with \eq{eq:dm2},
we first express the reduced density matrix 
in its Wigner representation,
\begin{equation}
    W(x,p) = 
    \frac{1}{\pi\hbar} \int dy \avg{x+y|\dm|x-y} e^{-2iyp/\hbar}.
\end{equation}
In terms of the Wigner function, \eq{eq:dm2} may be written
\begin{equation}
\label{eq:Wquantum}
	\partial_t{W}(x, p;\fm;t)  = \left( \LL_{cl} + \LL_q \right) W,
\end{equation}
where the evolution is described by two Liouvillian operators,
\begin{subequations}
\label{eq:Lops}
\begin{align}
\begin{split}
\label{eq:Lcl}
	\LL_{cl} &= - \frac{p}{M} \partial_x 
    + M \Omega^2 x \partial_p +
    \gamma \partial_p \cdot p + D \partial_p^2
    \\
    &\quad +   \left( e^{i\fm} - 1 \right)   \Gamma
    \left( 1 + \lambda x \right)^2, 
    \end{split} \\
    \begin{split}
    \label{eq:Lq}
	\LL_q &=   
	   \left( e^{i\fm} - 1 \right) 
	   \\ &\times
	   \bigg[   \frac{\gamma_1 \partial_p \cdot p + D_1 \partial_p^2}{2}
    +     \frac{\hbar^2}{4\pi M} \tunx \left( \tuno + \tunx x \right) \partial_x 
      \bigg] ,
      \end{split}
\end{align}
\end{subequations}
with the tunneling rate for the oscillator at $x=0$ given by
$\Gamma = \tuno^2 eV / 2\pi\hbar = 2D_1 \tuno^2 / \hbar^2 \tunx^2$.
The Liouvillian operator $\LL_{cl}$ 
describes the effectively classical evolution of the system: the first line
of \eq{eq:Lcl} corresponds to a classical Fokker-Planck equation
for the oscillator coupled to two effective equilibrium baths,
the environment and the junction;
the second line
describes tunneling as a classical Poisson process
characterized by a rate $\Gamma(t)$ that depends
on the instantaneous oscillator position $x(t)$.
In contrast, $\LL_q$ accounts for
quantum corrections to the effectively
classical evolution.
The $\fm$-dependent
terms involving $\gamma_1$ and $D_1$ in \eq{eq:Lq}
describe {\it conditional} damping and diffusion;
these terms represent back-action that is correlated with tunneling.  
Conditional back-action arises 
because each tunneling electron imparts a random
momentum kick to the oscillator, 
implying that the momentum kicks are correlated in time
with tunneling events,
and shows that
the back-action of the tunnel junction is {\it not}
fully described as an effective equilibrium bath.
This is discussed in detail in Section \ref{sec:NGC}.
Note that when we add the back-action terms in $\LL_q$ to
those in $\LL_{cl}$, we find that exactly {\it half} of the
total back-action is conditional (i.e. includes the factor $e^{i \fm}$).
The other half of the back-action is uncorrelated with tunneling, and
cannot be understood in terms of momentum
kicks imparted by tunneling electrons. 
We thus have the surprising conclusion that 
even during periods where no electrons tunnel, 
there is still back-action diffusion and damping.  
Heuristically, even if no electrons tunnel, we nonetheless
gain information about the oscillator and therefore
there must be back-action.
Finally, the remaining terms in \eq{eq:Lq} are also quantum in nature,
and arise from the difference between 
two tunneling processes 
involving absorption or emission of a phonon of energy $\hbar\Omega$.
In particular, the last term $\propto \tunx^2 x \partial_x W$
does not vanish when we trace over the oscillator degrees of freedom,
and thus represents a quantum correction to the 
average tunneling rate \cite{Mozyrsky02,ClerkGirvin04}
(cf. \eq{eq:Gx} below).
The same correction 
is obtained from a direct calculation
of the tunneling rate
using Fermi's golden rule.

\eq{eq:Wquantum} may be solved exactly
for the physical initial conditions of a thermal oscillator state.
Such a state is Gaussian,
and remains Gaussian under \eq{eq:Wquantum} for all times. 
We also take $m=0$ at time $t=0$, since this is when
we start counting tunneled electrons.
Thus, the Wigner function may be written in the form
\begin{equation}
	W(x,p;\fm;t) 
	= \frac{e^\phase 
	 e^{-\left[\frac{\pvar(x-\xx)^2+\xvar(p-\pp)^2-2\xpvar(x-\xx)(p-\pp)}
				{2(\xvar\pvar-\xpvar^2)} \right]}} {2\pi \sqrt{\xvar\pvar - \xpvar^2}} ,
\end{equation}
where we have scaled all quantities by the natural units of
the zero-point motion of the oscillator,
$\zx = \sqrt{\hbar/2M\Omega}$ 
and $\zp = \sqrt{M\hbar\Omega /2}$.
The state is
fully characterized at all times by its means,
$\xx$ and $\pp$,
its variances, $\xvar$ and $\pvar$,
its covariance, $\xpvar$, and
its normalization, $e^\phase$.
These six Gaussian parameters 
depend on both $\fm$ and $t$, and satisfy simple ordinary
differential equations which follow directly
from \eq{eq:Wquantum}.
First, the means satisfy
\begin{subequations}
\label{eq:avgs}
\begin{align}
	\begin{split}
	\partial_t{\xx}(\fm;t) 
	&=  \Omega \pp 
 	\\
	&+ \gamma_1 \left( e^{i\fm}-1 \right)
     	\left( \xx+\frac{1}{\lambda} \right)
    	\left( \frac{2T_1}{\hbar\Omega} \xvar - \frac{1}{2} \right),
 	\end{split}
	\label{eq:x} \\	
	\begin{split}
	 \partial_t {\pp}(\fm;t) &= 
  	 -\Omega \xx - \gamma \pp \\
	&+ \gamma_1\left( e^{i\fm}-1 \right)
    	\left[ \frac{2 T_1}{\hbar\Omega} \xpvar \left(\xx+\frac{1}{\lambda}\right) - \frac{\pp}{2} \right],
	\end{split}
  	\label{eq:p}
\end{align}
\end{subequations}
where we have again scaled the position and momentum 
by $\zx$ and $\zp$,
and defined the dimensionless coupling strength
$\lambda = \zx  \tunx / \tuno $.
The $\fm$-dependence of $\xx$ and $\pp$ encodes 
correlations between the 
oscillator motion and $m$.
For example, one can easily show that
the irreducible correlation between $x$
and the $n$th moment of $m$ is
$\redavg{x m^n} = (-i)^n \frac{\partial^n \xx}{\partial \fm^n}$.
Next, the variances and covariance (also scaled by $\zx$, $\zp$) satisfy
\begin{subequations}
\label{eq:vars}
\begin{align}
	\label{eq:xvar} 
	\begin{split}
	\partial_t {\var}_{x}(\fm;t) &= 2\Omega\xpvar 
	\\ & + \gamma_1\left( e^{i\fm}-1 \right)
    	\xvar \left( \frac{2 T_1}{\hbar\Omega} \xvar - 1 \right),
 	\end{split} \\
 	\label{eq:pvar}
   \begin{split}
   \partial_t{\var}_p(\fm;t) &=
  	 -2\Omega\xpvar 
	 \\& - 2\gamma_0 \bigg(\pvar-\frac{2\Tcoth}{\hbar\Omega} \bigg)
	- 2\gamma_1 \bigg(\pvar-\frac{2T_1}{\hbar\Omega} \bigg) 
	\\ &- \gamma_1\left( e^{i\fm} -1  \right)
    	\left( \pvar - \frac{2 T_1}{\hbar\Omega} \left( 1 + \xpvar^2\right) \right),
   \end{split} \\
 \begin{split}
  \partial_t{\var}_{xp}(\fm;t) &= 
  	 \Omega \left( \pvar-\xvar \right) -\gamma\xpvar  
	 \\ &
	 + \gamma_1\left( e^{i\fm}-1 \right)
    	\xpvar \left( \frac{2 T_1}{\hbar\Omega} \xvar - 1 \right),
	\label{eq:xpvar} 
  \end{split} 
\end{align}
\end{subequations}
where $\Tcoth = (\hbar\Omega/2) \coth{\left(\hbar\Omega/2T_0\right)}$.
Again, the $\fm$-dependence of these parameters
describes correlations
between $x^2$, $p^2$ or $xp$ and moments of $m$.
Finally, the parameter $\phase$ satisfies
\begin{equation}
	\partial_t{\phase} =   \left( e^{i\fm}-1 \right) \\
   	\left[ \Gamma \left( 1 + 2 \lambda\xx + \lambda^2(\xx^2 + \xvar)
     	\right) - \frac{\gamma_1}{2} \right] ,
\label{eq:phase} 
\end{equation}
and is directly connected to the FCS
as discussed in the next section.

Eqs.~(\ref{eq:avgs}--\ref{eq:phase})
have simple analytic solutions
in the limit of long times, 
and may be solved numerically for all times to
arbitrary precision.
Before using the equations to study
the FCS and conditional evolution, 
we emphasize an important difference
from previous treatments of conditional
evolution in NEMS: 
the evolution of the variances is conditional,
as seen directly
from the $\fm$-dependent terms in Eqs.~(\ref{eq:vars}).
This is in stark contrast to 
the standard treatment 
where the variances evolve independently of tunneling \cite{Hopkins03,Ruskov04}.
This is partly due to the conditional back-action
diffusion in $\LL_q$ discussed above (cf. \eq{eq:Lq}),
which implies that momentum fluctuations of the oscillator
are correlated with fluctuations in $m$ and 
leads to the conditional terms in \eq{eq:pvar}.
However,
we also find conditional terms in \eq{eq:xvar}
that arise from the {\it classical} part of
\eq{eq:Wquantum} described by $\LL_{cl}$.
This is because we start with
the linear $x$-dependence
of the tunneling {\it amplitude} in \eq{eq:Ht}, and it 
follows that the tunneling {\it rate}
has both  linear and  quadratic $x$-dependence:
\begin{equation}
\label{eq:Gx}
	\Gx = \Gamma 
	\left[ 1 +  2 \lambda x + \lambda^2 x^2  \right] - \frac{\gamma_1}{2}.
\end{equation}
Standard treatments of conditional evolution neglect the quadratic dependence, 
which in our case is inconsistent with the starting Hamiltonian
\footnote{Including the quadratic $x$-dependence in the amplitude leads to
a renormalization of $\lambda$ as well as 
two-phonon absorption/emission during a tunneling event;
both are unimportant in the limit of weak coupling.}.
We stress that the conditional (i.e.~$\fm$-dependent) 
and unconditional (i.e.~$\fm$-independent) 
terms in Eqs.~(\ref{eq:vars}) appear at the same order
in the coupling strength $\lambda$;
there is no a priori reason to keep one effect and not the other.
The results presented below are contingent on the conditional
evolution of the variances.


\section{Full counting statistics}

\begin{figure}[htb]
\centering
	{\includegraphics[width=\bigfigsize]{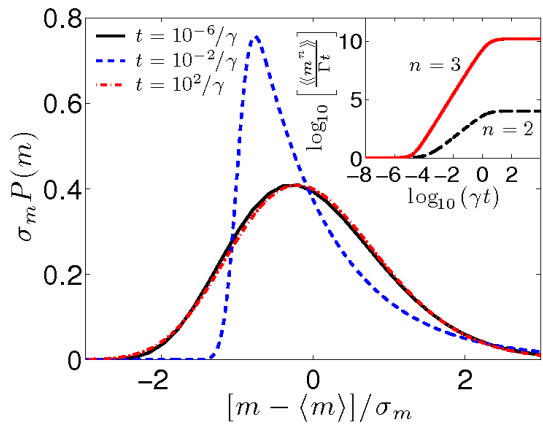}}
\caption{(Color online).  Main: $P(m;t)$ at three times.
We have shifted $m$ 
by its average $\avg{m(t)}$ and scaled by its standard deviation 
$\sigma_m(t)$.
Inset: oscillator enhancement of 
the variance (black dashed) and skewness (red solid) of $P(m;t)$ 
versus time.
We have taken 
$T_0=T_1=10^4 \hbar\Omega,\lambda=10^{-3},\tau_0=0.2,\gamma_0=10^{-5}\Omega$.
These values are based on the experiment in Ref.~\onlinecite{Flowers-Jacobs07},
except that we decreased the intrinsic tunneling strength $\tau_0$ to operate in
the tunneling regime and increased
the coupling strength $\lambda$ to clearly illustrate our results.
For these values, the fractional change in the average current due to the oscillator is 
$\lambda^2 \redavgSm{x^2} = 0.02$.}
\label{fig:fcs}
\end{figure}

It follows directly from Eq.~(\ref{eq:fcs}) 
that the generating function for the FCS is given by the
Gaussian parameter $\phi$ 
via
$\charFunc(\fm;t) = e^{\phi(\fm;t)}$.
From \eq{eq:phase} we see that if the average and variance
of the oscillator were simply constants,
then tunneling electrons would obey Poisson statistics with an effective
tunneling rate given by $\langle \Gx \rangle$,
obtained from \eq{eq:Gx}.
However, 
the oscillator position is correlated with tunneling
electrons; this correlation enters
\eq{eq:phase} through the 
$\fm$-dependence of $\xx$ and $\xvar$ and leads
to deviations from Poisson statistics.
From \twoEq{eq:fcs}{eq:phase} we obtain 
$P(m;t)$, shown at several times in \fig{fig:fcs}.
Even for weak coupling, i.e. $\lambda^2 \redavg{x^2} \ll  1$, 
the oscillator can have a dramatic effect on $P(m;t)$
at what we call intermediate times,
$\tstar \lesssim t \lesssim 1/\gamma$, causing it to become
highly non-Gaussian.  Here,
\begin{equation}
\label{eq:tstar}
	\tstar \sim \frac{1}{\Gamma} \left( \frac{\hbar\Omega}{\lambda^2 T}\right)^2,
\end{equation}
and $T = D/ M \gamma$ is the net effective temperature of the oscillator
due to both the tunnel junction and the thermal environment. 
We emphasize that in the relevant limit
of a high-$Q$ oscillator, 
the timescale $1/\gamma$ is much larger than
$1/\Gamma$
and thus many electrons have tunneled
even for intermediate times.

The significant modification of the FCS is due to the seemingly weak dependence
of the current on $x^2$ (cf.~Eq.~(\ref{eq:Gx})).  To see this, it is useful to consider 
the first few cumulants  of $m$.
From Eqs.~(\ref{eq:fcs}) and (\ref{eq:phase}), 
these satisfy ($x$ is again scaled by $\zx$)
\begin{subequations}
\label{eq:moments}
\begin{align}
	\partial_t{\redavg{m^2}} &= \partial_t{\avg{m}} + 2\Gamma \left[
		2\lambda \redavg{xm} + \lambda^2 \redavg{x^2 m} \right],
	\label{eq:mVar} \\
	\partial_t{\redavg{m^3}} &= \partial_t{\avg{m}} + 3\Gamma \bigg\{
		2\lambda \left[ \redavg{xm} + \redavg{xm^2} \right] 
		\label{eq:m3} \\
		&+  \lambda^2 \left[ \redavg{x^2 m} + \redavg{x^2 m^2}
		+  \redavg{x m}^2 \right]  \bigg\} , \nonumber
\end{align}
\end{subequations}
where all of correlations depend on $t$.
The first term in each equation 
corresponds to Poisson
statistics, in which all cumulants would be equal to $\avg{m}$.
Correlations of $x$  and $x^2$ with
$m$ emerge naturally in the cumulants
due to the $x$-dependence of the tunneling rate in
\eq{eq:Gx}, because $m(t) = \int_0^t dt' \tilde{\Gamma}[x(t')]$.
As indicated in Eqs.~(\ref{eq:moments}), this allows
oscillator fluctuations to affect the variance and skewness of $m$; 
since $x$ and $x^2$ are positively correlated with $m$,
the cumulants will be increased by these correlations.
Similar correlations 
appear in the higher moments.
These correlations can strongly affect the cumulants
due to the slow decay of energy fluctuations in the oscillator,
as we now discuss.


Consider \eq{eq:mVar}.
From the $x^2$ term in $\Gx$, 
fluctuations in $x^2$ will lead to fluctuations in 
$m$.
Thus the last term in \eq{eq:mVar} leads to
a term
$2\Gamma^2  \lambda^4 \int_0^t \int_0^t dt_1 dt_2 \redavg{x^2(t_1) x^2(t_2)}$
in the variance.
The factor $\lambda^4$ is small due to weak coupling; however, the
$x^2$ autocorrelation in the integrand is
proportional to an energy autocorrelation (up to insignificant rapidly
oscillating terms).
This contribution initially scales as $(T/\hbar\Omega)^2$ and decays  on the very slow timescale of the 
oscillator ring-down time, $1/\gamma$.  
Thus, long-lived energy fluctuations
in the high-$Q$ oscillator allow its influence 
to build up, eventually overcoming the 
weak coupling strength
and dominating the FCS.
This enhancement occurs when
the last term in \eq{eq:mVar} dominates the first,
requiring $\Gamma (\lambda^2 T/\hbar\Omega)^2 / \gamma \gg 1$.
This condition can be satisfied even when the oscillator contribution to the average current 
$ e \langle \Gx \rangle$ is small, as
the ratio $\Gamma/\gamma$ is typically large
(e.g. $\Gamma/\gamma\sim 10^8$ in Ref.~\onlinecite{Flowers-Jacobs07}).
Further, this same condition ensures $\tstar \ll 1/\gamma$ from \eq{eq:tstar}, 
resulting in non-Gaussian FCS over a wide range of times.

If the condition for enhancement is met, then
the effect is even greater for higher cumulants.
For example, \eq{eq:m3}
contains a term proportional to $\redavg{x^2(t) m^2(t)}$.
This leads to an oscillator-dependent term in the skewness similar to that in
the variance, with an
additional factor of $m$ resulting in
an extra factor of $\Gamma \lambda^2 x^2$ and
an extra time integral.  
We obtain a three-time $x^2$ autocorrelation
which initially scales as $(T/\hbar\Omega)^3$ and decays
on the timescale $1/\gamma$, compensating for the extra factor
of weak coupling.
In general, we find that the maximum enhancement for the $n$th cumulant is roughly
\begin{equation}
\label{eq:nthCumulant}
	\redavg{m^n(t)} \sim \left( \Gamma \lambda ^2 T t  /\hbar\Omega   \right)^n
\end{equation}
for times $\tstar \ll t \ll 1/\gamma$.
This can be seen directly from \eq{eq:Gx} by assuming that fluctuations
in $m$ are dominated by $x^2$ fluctuations for this range of times.

\fig{fig:fcs} shows that
$P(m;t)$ is skewed only for intermediate times
$\tstar \lesssim t \lesssim 1/\gamma$; 
for short and long times the distribution is nearly Gaussian.  The enhancement of cumulants compared to their Poisson values (i.e. $\redavg{m^n(t)}=\Gamma t$ with no oscillator) is shown in the inset of \fig{fig:fcs}.
For short times, the effects of the weakly coupled oscillator 
have not yet built up and we
obtain the Poisson statistics of the
uncoupled tunnel junction.  
For long times $t \gg 1/\gamma$, the contribution 
to $\redavg{ m^n }$ 
from $x^2$ fluctuations simply scales as $t$
(and not as $t^n$), as $t$ is now much longer than the lifetime of a typical
oscillator energy fluctuation.  
For long times the oscillator still enhances 
$\redavg{ m^n }$ by a factor
$(\Gamma/\gamma)^{n-1} (\lambda^2 T/\hbar\Omega)^n$ over the Poisson
value $\Gamma t$, but
since each cumulant is proportional to $t$, 
$P(m;t)$ tends to a Gaussian
\footnote{The $n$th cumulant
{\it relative} to the width is $\redavg{m^n} / (\Delta m)^n$, which
vanishes for long times.
This satisfies the central limit theorem.}.
To estimate
the timescale $\tstar$ for the buildup of enhanced cumulants, note that
significant enhancement will occur when
the oscillator contribution to the variance in \eq{eq:mVar} is larger than the 
Poisson contribution.
From \eq{eq:nthCumulant} this requires 
$\left( \Gamma \lambda^2 T t /\hbar\Omega \right)^2 > \Gamma t$,
which yields \eq{eq:tstar} for the timescale $\tstar$.

In the range of times where the FCS is strongly influenced by
the oscillator, $P(m;t)$ is directly related to $P(x)$.
For a thermal oscillator at temperature $T$, we have
$P(x) = \sqrt{\frac{\hbar\Omega}{4\pi T} } e^{- \hbar\Omega x^2 / 4T}$
with $x$ in units of $\zx$.
Assuming that fluctuations of $x^2$ are the dominant source of large $m$ fluctuations, and using \eq{eq:Gx}, we obtain
\begin{equation}
	P(m;t) 
	\propto \exp{ \left[ -\frac{\hbar\Omega m}{ 4\Gamma \lambda^2 T t}    \right] }
\end{equation}
for $m \gg \Gamma t$.
This estimate describes the tail of 
$P(m;t)$ very well for times
$\tstar \ll t \ll 1/\gamma$.

\section{Conditional evolution}

\begin{figure}[htb]
\centering
	\includegraphics[width=\smallfigsize]{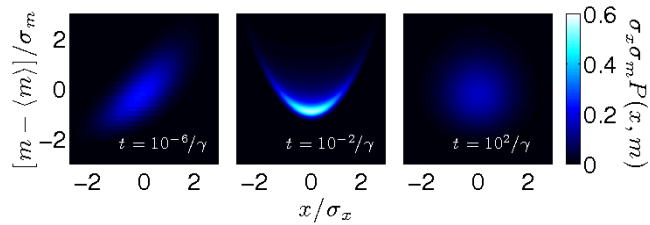}
\caption{(Color online).  Joint distribution $P(x,m;t)$ calculated for 
the same times and parameters as in \fig{fig:fcs}.
We have shifted $m$ by its average $\avg{m(t)}$ and scaled
$x$ and $m$ by their
standard deviations, $\sigma_x(t)$ and $\sigma_m(t)$.}
\label{fig:Pxm}
\end{figure}

The effects of the oscillator on the FCS are the result of
correlations between $x^2$ and $m$;
we can thus gain further insight by studying conditional dynamics.
The joint distribution
$P(x,m;t) = \avg{x | \dm(m;t) |x }$ is shown  in \fig{fig:Pxm}.
Consistent with the FCS, for short and long times we see only small correlations
between $x$ and $m$.
$P(x,m;t)$ is
most striking at times $\tstar \lesssim t \lesssim 1/\gamma$
due to correlations between $x^2$ and $m$.

\begin{figure}[htb]
\centering
	\includegraphics[width=\smallfigsize]{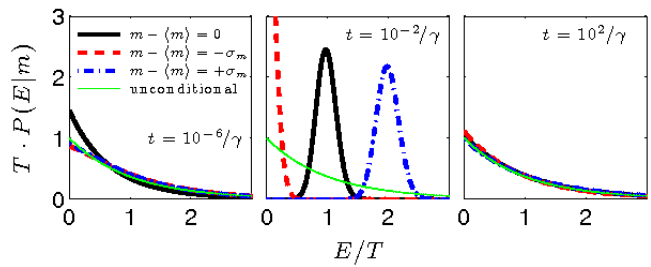}
\caption{(Color online). Conditional distributions $P(E|m;t)$ for the same times and 
parameters  as in \fig{fig:fcs}.  The green (thin solid) line shows the unconditional
distribution (average over $m$).}
\label{fig:cond}
\end{figure}

Eqs.~(\ref{eq:avgs}--\ref{eq:phase})
 may also be used to find 
the conditional energy
distribution, $P(E|m;t)$---given a particular measurement 
history and value of $m(t)$, what is the oscillator's energy distribution?
In \fig{fig:cond} we
see that 
for  $\tstar \ll t \ll 1/\gamma$,
the conditional energy distributions are highly non-thermal and 
localized at the energy required to produce the given value of $m$
from \eq{eq:Gx}, with
width given roughly by $T$.
The ability to obtain information about the oscillator's energy 
distribution using a 
weakly coupled detector is somewhat surprising, and is another result
of long-lived energy fluctuations in the oscillator.

\section{Non-Gaussian corrections to the effective bath model}
\label{sec:NGC}

The effects discussed so far are 
captured by the effectively classical Liouvillian operator $\LL_{cl}$
of \eq{eq:Wquantum}.
Neglecting the quantum corrections  results in
an ``effective bath" model, where the back-action effects of the tunnel
junction are treated as arising from a 
second thermal bath coupled to the oscillator,
and the oscillator is treated as a classical variable which sets 
the instantaneous tunnelling rate.  
However, 
the conditional back-action damping and diffusion terms in
\eq{eq:Lq} lead to  
non-Gaussian correlations between the junction current and back-action
force operators $\hat{I}$ and $\hat{F}$ that
are {\it not} captured by the effective bath model.
These arise 
because even though
tunneling is stochastic and imparts random momentum 
kicks to the oscillator, 
each momentum kick occurs
{\it at the same time} that an electron tunnels.  
This is completely missed in the effective bath model, as it 
treats the 
junction as a thermal noise source 
independent of individual tunneling events.  For example,
using Eqs.~(\ref{eq:vars}) to calculate $\redavg{x^2(t) m(t)}$ in the long
time limit, we find
an enhancement compared to the effective bath model:
\begin{equation}
	\Delta  \redavg{x^2(t) m(t)}_Q =  \frac{\gamma_1 T_1}{\gamma\hbar\Omega}
	\quad\quad\text{($t\rightarrow\infty$)}.
\label{eq:Qcorrelation}
\end{equation} 
This implies the existence of non-Gaussian correlations
between the the current and back-action force.
A direct quantum calculation of the non-Gaussian correlator
$\redavgSm{ \hat{F}(t_1) \hat{F}(t_2) \hat{I}(t_3) }$
using Keldysh path integrals following Ref.~\onlinecite{Levitov04} 
leads to the same non-Gaussian correction given in \eq{eq:Qcorrelation}.

The non-Gaussian correlations 
may be understood in terms of a simple model of quantum back-action.
We describe the oscillator-independent tunneling current
as a sequence of $\delta$-functions,
		$I(t) = e \sum_{n=0}^\infty \delta(t-t_n)$, 
where the intervals between the $t_n$ are exponentially distributed.
The back-action force of the junction is then taken to be
	$F(t) = \sum_{n=0}^\infty \xi_n  \delta(t-t_n)$,
where $\xi_n$ is a zero-mean random variable
describing the impulse imparted to the oscillator by the $n$th
electron.  The same sequence of times $\{t_n\}$ appears in both
$I(t)$ and $F(t)$, reflecting the fact
that back-action arises from the action of individual tunneling electrons.
If we then take $\avg{\xi_m \xi_n} = (\hbar \tunx / \tuno)^2 \delta_{mn}$, our
simple model reproduces the non-Gaussian correlations obtained from 
Eqs.~(\ref{eq:avgs}--\ref{eq:phase}); 
we also obtain the expected
back-action diffusion constant $D_1$.  
From the size of $\xi_n$  we see that the typical
momentum kick imparted by a single tunneling electron  
is given by $\Delta p \sim \hbar \tunx / \tuno$, and not by the Fermi momentum.
This value for $\Delta p$ is consistent with the Heisenberg uncertainty
principle, since the sensitivity of a position measurement
scales as $\Delta x \sim \tuno / \tunx$.
We thus have a simple picture for the 
source of the conditional part of quantum back-action: 
it arises from tunneling electrons imparting
random momentum kicks of size set by the uncertainty principle.
Again we stress that this picture only accounts
for the {\it conditional} half of the back-action
damping and diffusion; the other
half is completely uncorrelated with tunneling electrons
(cf. \eq{eq:Lq} and the discussion thereafter).
We also note that one can derive the
conditional back-action terms in \eq{eq:Lq} directly
from this simple model, from a corresponding classical master
equation in which each tunneling event is associated with a
random momentum kick.

\begin{figure}[htb]
\centering
	\includegraphics[width=\bigfigsize]{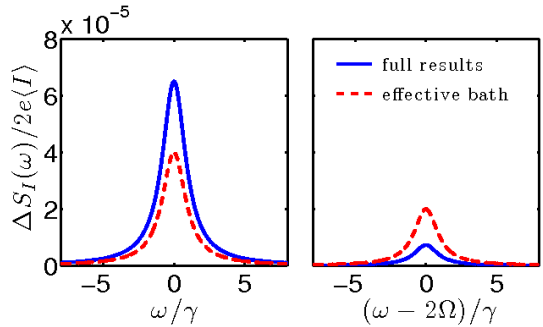}
\caption{(Color online). Contribution to the 
current noise spectrum near $\omega=0$ (left)
and $\omega=2\Omega$ (right) 
from to the correlation $\redavg{x^2m}$.  The
full calculation including non-Gaussian corrections (blue solid)
are compared to the results from the effective bath model (red dashed).
The contributions are normalized
by the frequency-independent shot noise background.
We have taken $T_0 = \hbar\Omega/2$, $T_1 = 100 \hbar\Omega$, $\lambda=0.01$,
$\tau_0=0.2$ and $\gamma_0 = 10^{-3} \Omega$ in order to approach the 
limit where the non-Gaussian signatures are maximized.
}
\label{fig:noise}
\end{figure}

The non-Gaussian correlations discussed above can in principle be detected 
via the finite-frequency current noise in the tunnel junction, $S_I(\omega)$.
This may be found from the time dependence
of $\redavg{m^2}$ using the 
MacDonald formula \cite{MacDonald48}, 
\begin{equation}
	S_I(\omega) = 2e^2 \omega \int_0^\infty dt \sin{(\omega t)}
	\partial_t{ \redavg{m^2(t)}}.
\end{equation}
Note that the frequency-dependent current noise
is obtained from
the {\it particle} current fluctuations only.
In the single junction,
tunneling is non-resonant and
there is no place for charge to build up in the system, so
displacement currents may be safely neglected
\cite{Blanter00}.
The time derivative of $\redavg{m^2}$ is given in \eq{eq:mVar},
which shows that we need the full, time-dependent
correlations $\redavg{x(t)m(t)}$ and $\redavg{x^2(t)m(t)}$
to calculate the current noise.
These correlations are calculated simply by taking the $\fm$-derivative
of Eqs.~(\ref{eq:avgs}--\ref{eq:phase}).
The resulting equations are readily solved exactly
for the physical initial conditions in which the 
oscillator is equilibrated
with both the environment and the tunnel junction, i.e. 
$\redavg{x^2} = \redavg{p^2} = 2T$ in our units.
The correlation $\redavg{xm}$ leads to a peak at $\omega=\Omega$
in the noise that is very accurately captured by
the effective bath model.
However, the correlation $\redavg{x^2m}$ leads
to peaks at $\omega=0$ and $\omega=2\Omega$
that show signatures of the non-Gaussian correlations.
This is especially true in the
limit $\gamma_0 / \gamma_1 \gg T_1 / T_0 \gg 1$,
where the non-Gaussian correlations lead
to a doubling of the current noise peak at $\omega=0$, and completely suppress
the peak at $\omega=2\Omega$ as shown in \fig{fig:noise}.
This limit requires a back-action damping rate much smaller than the
intrinsic damping from the environment, and
a back-action temperature much greater than
the temperature of the environment.
The first of these conditions is natural in experiments and the second
has been achieved.
Note that we still require the intrinsic damping of the oscillator to be small.

Measurements of $S_I(\omega)$ 
could thus be used to distinguish the tunnel junction's back-action
on the oscillator
from the effects of a Gaussian uncorrelated noise source 
supplied by an equilibrium bath.
However, as seen in \fig{fig:noise}, in the same limit where the signatures
are relatively large, the peak heights themselves are very small compared
to the frequency-independent shot noise background.
For this reason, detecting these signatures in the current noise would pose
a formidable challenge.  Further thought will be devoted to more efficient
strategies to detect the non-Gaussian correlations we have identified.

\section{Conclusions}

We have studied the statistics of the experimentally relevant NEMS
of a tunnel junction coupled to a mechanical oscillator.
We have shown that even if the coupling is very weak, 
long-lived energy fluctuations
in the oscillator allow it dominate the FCS.
The oscillator-induced enhancement of the 
third moment of the FCS
could be observed up to 
measurement times
near $1/\gamma$, well within reach of current 
experiments.  
We have also shown that the effective bath model is
not sufficient to fully describe the effects of the tunnel junction
on the oscillator.
Half of the back-action is conditional as a result of the random
momentum kicks imparted to the oscillator by tunneling electrons,
and this leads to non-Gaussian correlations with 
signatures in the finite frequency current noise.


This work was supported by NSERC and CIFAR.

\bibliography{../AllRefs}

\begin{thebibliography}{20}
\expandafter\ifx\csname natexlab\endcsname\relax\def\natexlab#1{#1}\fi
\expandafter\ifx\csname bibnamefont\endcsname\relax
  \def\bibnamefont#1{#1}\fi
\expandafter\ifx\csname bibfnamefont\endcsname\relax
  \def\bibfnamefont#1{#1}\fi
\expandafter\ifx\csname citenamefont\endcsname\relax
  \def\citenamefont#1{#1}\fi
\expandafter\ifx\csname url\endcsname\relax
  \def\url#1{\texttt{#1}}\fi
\expandafter\ifx\csname urlprefix\endcsname\relax\def\urlprefix{URL }\fi
\providecommand{\bibinfo}[2]{#2}
\providecommand{\eprint}[2][]{\url{#2}}

\bibitem[{\citenamefont{Flowers-Jacobs
  et~al.}(2007)\citenamefont{Flowers-Jacobs, Schmidt, and
  Lehnert}}]{Flowers-Jacobs07}
\bibinfo{author}{\bibfnamefont{N.~E.} \bibnamefont{Flowers-Jacobs}},
  \bibinfo{author}{\bibfnamefont{D.~R.} \bibnamefont{Schmidt}},
  \bibnamefont{and} \bibinfo{author}{\bibfnamefont{K.~W.}
  \bibnamefont{Lehnert}}, \bibinfo{journal}{Phys. Rev. Lett.}
  \textbf{\bibinfo{volume}{98}}, \bibinfo{eid}{096804} (\bibinfo{year}{2007}).

\bibitem[{\citenamefont{Poggio et~al.}(2008)\citenamefont{Poggio, Jura, Degen,
  Topinka, Mamin, Goldhaber-Gordon, and Rugar}}]{Poggio08}
\bibinfo{author}{\bibfnamefont{M.}~\bibnamefont{Poggio}},
  \bibinfo{author}{\bibfnamefont{M.~P.} \bibnamefont{Jura}},
  \bibinfo{author}{\bibfnamefont{C.~L.} \bibnamefont{Degen}},
  \bibinfo{author}{\bibfnamefont{M.~A.} \bibnamefont{Topinka}},
  \bibinfo{author}{\bibfnamefont{H.~J.} \bibnamefont{Mamin}},
  \bibinfo{author}{\bibfnamefont{D.}~\bibnamefont{Goldhaber-Gordon}},
  \bibnamefont{and} \bibinfo{author}{\bibfnamefont{D.}~\bibnamefont{Rugar}},
  \bibinfo{journal}{Nat. Phys.} \textbf{\bibinfo{volume}{4}},
  \bibinfo{pages}{635} (\bibinfo{year}{2008}).

\bibitem[{\citenamefont{Knobel and Cleland}(2003)}]{Knobel03}
\bibinfo{author}{\bibfnamefont{R.~G.} \bibnamefont{Knobel}} \bibnamefont{and}
  \bibinfo{author}{\bibfnamefont{A.~N.} \bibnamefont{Cleland}},
  \bibinfo{journal}{Nature} \textbf{\bibinfo{volume}{424}},
  \bibinfo{pages}{291} (\bibinfo{year}{2003}).

\bibitem[{\citenamefont{Naik et~al.}(2006)\citenamefont{Naik, Buu, LaHaye,
  Armour, Clerk, Blencowe, and Schwab}}]{Naik06}
\bibinfo{author}{\bibfnamefont{A.}~\bibnamefont{Naik}},
  \bibinfo{author}{\bibfnamefont{O.}~\bibnamefont{Buu}},
  \bibinfo{author}{\bibfnamefont{M.~D.} \bibnamefont{LaHaye}},
  \bibinfo{author}{\bibfnamefont{A.~D.} \bibnamefont{Armour}},
  \bibinfo{author}{\bibfnamefont{A.~A.} \bibnamefont{Clerk}},
  \bibinfo{author}{\bibfnamefont{M.~P.} \bibnamefont{Blencowe}},
  \bibnamefont{and} \bibinfo{author}{\bibfnamefont{K.~C.}
  \bibnamefont{Schwab}}, \bibinfo{journal}{Nature}
  \textbf{\bibinfo{volume}{443}}, \bibinfo{pages}{193} (\bibinfo{year}{2006}).

\bibitem[{\citenamefont{Armour}(2004)}]{Armour04}
\bibinfo{author}{\bibfnamefont{A.~D.} \bibnamefont{Armour}},
  \bibinfo{journal}{Phys. Rev. B} \textbf{\bibinfo{volume}{70}},
  \bibinfo{pages}{165315} (\bibinfo{year}{2004}).

\bibitem[{\citenamefont{Zhang and Blencowe}(2002)}]{ZhangBlencowe02}
\bibinfo{author}{\bibfnamefont{Y.}~\bibnamefont{Zhang}} \bibnamefont{and}
  \bibinfo{author}{\bibfnamefont{M.~P.} \bibnamefont{Blencowe}},
  \bibinfo{journal}{J. Appl. Phys.} \textbf{\bibinfo{volume}{91}},
  \bibinfo{pages}{4249} (\bibinfo{year}{2002}).

\bibitem[{\citenamefont{Usmani et~al.}(2007)\citenamefont{Usmani, Blanter, and
  Nazarov}}]{Usmani07}
\bibinfo{author}{\bibfnamefont{O.}~\bibnamefont{Usmani}},
  \bibinfo{author}{\bibfnamefont{Y.~M.} \bibnamefont{Blanter}},
  \bibnamefont{and} \bibinfo{author}{\bibfnamefont{Y.~V.}
  \bibnamefont{Nazarov}}, \bibinfo{journal}{Phys. Rev. B}
  \textbf{\bibinfo{volume}{75}}, \bibinfo{eid}{195312} (\bibinfo{year}{2007}).

\bibitem[{\citenamefont{Clerk and Girvin}(2004)}]{ClerkGirvin04}
\bibinfo{author}{\bibfnamefont{A.~A.} \bibnamefont{Clerk}} \bibnamefont{and}
  \bibinfo{author}{\bibfnamefont{S.~M.} \bibnamefont{Girvin}},
  \bibinfo{journal}{Phys. Rev. B} \textbf{\bibinfo{volume}{70}},
  \bibinfo{pages}{121303(R)} (\bibinfo{year}{2004}).

\bibitem[{\citenamefont{Mozyrsky and Martin}(2002)}]{Mozyrsky02}
\bibinfo{author}{\bibfnamefont{D.}~\bibnamefont{Mozyrsky}} \bibnamefont{and}
  \bibinfo{author}{\bibfnamefont{I.}~\bibnamefont{Martin}},
  \bibinfo{journal}{Phys. Rev. Lett.} \textbf{\bibinfo{volume}{89}},
  \bibinfo{pages}{018301} (\bibinfo{year}{2002}).

\bibitem[{\citenamefont{Doiron et~al.}(2008)\citenamefont{Doiron, Trauzettel,
  and Bruder}}]{Doiron08}
\bibinfo{author}{\bibfnamefont{C.~B.} \bibnamefont{Doiron}},
  \bibinfo{author}{\bibfnamefont{B.}~\bibnamefont{Trauzettel}},
  \bibnamefont{and} \bibinfo{author}{\bibfnamefont{C.}~\bibnamefont{Bruder}},
  \bibinfo{journal}{Phys. Rev. Lett.} \textbf{\bibinfo{volume}{100}},
  \bibinfo{eid}{027202} (\bibinfo{year}{2008}).

\bibitem[{\citenamefont{Wabnig et~al.}(2005)\citenamefont{Wabnig, Khomitsky,
  Rammer, and Shelankov}}]{Wabnig05}
\bibinfo{author}{\bibfnamefont{J.}~\bibnamefont{Wabnig}},
  \bibinfo{author}{\bibfnamefont{D.~V.} \bibnamefont{Khomitsky}},
  \bibinfo{author}{\bibfnamefont{J.}~\bibnamefont{Rammer}}, \bibnamefont{and}
  \bibinfo{author}{\bibfnamefont{A.~L.} \bibnamefont{Shelankov}},
  \bibinfo{journal}{Phys. Rev. B} \textbf{\bibinfo{volume}{72}},
  \bibinfo{eid}{165347} (\bibinfo{year}{2005}).

\bibitem[{\citenamefont{Levitov and Reznikov}(2004)}]{Levitov04}
\bibinfo{author}{\bibfnamefont{L.~S.} \bibnamefont{Levitov}} \bibnamefont{and}
  \bibinfo{author}{\bibfnamefont{M.}~\bibnamefont{Reznikov}},
  \bibinfo{journal}{Phys. Rev. B} \textbf{\bibinfo{volume}{70}},
  \bibinfo{eid}{115305} (\bibinfo{year}{2004}).

\bibitem[{\citenamefont{Bomze et~al.}(2005)\citenamefont{Bomze, Gershon,
  Shovkun, Levitov, and Reznikov}}]{Bomze05}
\bibinfo{author}{\bibfnamefont{Y.}~\bibnamefont{Bomze}},
  \bibinfo{author}{\bibfnamefont{G.}~\bibnamefont{Gershon}},
  \bibinfo{author}{\bibfnamefont{D.}~\bibnamefont{Shovkun}},
  \bibinfo{author}{\bibfnamefont{L.~S.} \bibnamefont{Levitov}},
  \bibnamefont{and} \bibinfo{author}{\bibfnamefont{M.}~\bibnamefont{Reznikov}},
  \bibinfo{journal}{Phys. Rev. Lett.} \textbf{\bibinfo{volume}{95}},
  \bibinfo{eid}{176601} (\bibinfo{year}{2005});
  \bibinfo{author}{\bibfnamefont{B.}~\bibnamefont{Reulet}},
  \bibinfo{author}{\bibfnamefont{J.}~\bibnamefont{Senzier}}, \bibnamefont{and}
  \bibinfo{author}{\bibfnamefont{D.~E.} \bibnamefont{Prober}},
  \bibinfo{journal}{Phys. Rev. Lett.} \textbf{\bibinfo{volume}{91}},
  \bibinfo{pages}{196601} (\bibinfo{year}{2003}).


\bibitem[{\citenamefont{Pistolesi}(2004)}]{Pistolesi04}
\bibinfo{author}{\bibfnamefont{F.}~\bibnamefont{Pistolesi}},
  \bibinfo{journal}{Phys. Rev. B} \textbf{\bibinfo{volume}{69}},
  \bibinfo{pages}{245409} (\bibinfo{year}{2004}).

\bibitem[{\citenamefont{C.~Flindt and Jauho}(2005)}]{Flindt05}
\bibinfo{author}{\bibfnamefont{T.~N.} \bibnamefont{C.~Flindt}}
  \bibnamefont{and} \bibinfo{author}{\bibfnamefont{A.-P.} \bibnamefont{Jauho}},
  \bibinfo{journal}{Europhys. Lett.} \textbf{\bibinfo{volume}{69}},
  \bibinfo{pages}{475} (\bibinfo{year}{2005}).

\bibitem[{\citenamefont{Ruskov et~al.}(2005)\citenamefont{Ruskov, Schwab, and
  Korotkov}}]{Ruskov04}
\bibinfo{author}{\bibfnamefont{R.}~\bibnamefont{Ruskov}},
  \bibinfo{author}{\bibfnamefont{K.}~\bibnamefont{Schwab}}, \bibnamefont{and}
  \bibinfo{author}{\bibfnamefont{A.~N.} \bibnamefont{Korotkov}},
  \bibinfo{journal}{Phys. Rev. B} \textbf{\bibinfo{volume}{71}},
  \bibinfo{eid}{235407} (\bibinfo{year}{2005}).

\bibitem[{\citenamefont{Hopkins et~al.}(2003)\citenamefont{Hopkins, Jacobs,
  Habib, and Schwab}}]{Hopkins03}
\bibinfo{author}{\bibfnamefont{A.}~\bibnamefont{Hopkins}},
  \bibinfo{author}{\bibfnamefont{K.}~\bibnamefont{Jacobs}},
  \bibinfo{author}{\bibfnamefont{S.}~\bibnamefont{Habib}}, \bibnamefont{and}
  \bibinfo{author}{\bibfnamefont{K.}~\bibnamefont{Schwab}},
  \bibinfo{journal}{Phys. Rev. B} \textbf{\bibinfo{volume}{68}},
  \bibinfo{pages}{235328} (\bibinfo{year}{2003}).

\bibitem[{\citenamefont{Caldeira and Leggett}(1983)}]{Caldeira83}
\bibinfo{author}{\bibfnamefont{A.~O.} \bibnamefont{Caldeira}} \bibnamefont{and}
  \bibinfo{author}{\bibfnamefont{A.~J.} \bibnamefont{Leggett}},
  \bibinfo{journal}{Ann. Phys. (N.Y.)} \textbf{\bibinfo{volume}{149}},
  \bibinfo{pages}{374} (\bibinfo{year}{1983}).

\bibitem[{\citenamefont{MacDonald}(1948)}]{MacDonald48}
\bibinfo{author}{\bibfnamefont{D.~K.~C.} \bibnamefont{MacDonald}},
  \bibinfo{journal}{Rep. Prog. Phys.} \textbf{\bibinfo{volume}{12}},
  \bibinfo{pages}{56} (\bibinfo{year}{1948}).

\bibitem[{\citenamefont{Blanter and B{\"{u}}ttiker}(2000)}]{Blanter00}
\bibinfo{author}{\bibfnamefont{Y.~M.} \bibnamefont{Blanter}} \bibnamefont{and}
  \bibinfo{author}{\bibfnamefont{M.}~\bibnamefont{B{\"{u}}ttiker}},
  \bibinfo{journal}{Phys. Rep.} \textbf{\bibinfo{volume}{336}},
  \bibinfo{pages}{1} (\bibinfo{year}{2000}).

\end{thebibliography}
\bibliographystyle{apsrev}

\end{document}